# Correlating on-the-fly Electrical and Optical Skyrmion Readout

**Authors**

Grischa Beneke[1], Kilian Leutner[1], Nikhil Vijayan[1,2], Fabian Kammerbauer[1], Duc Minh Tran[1], Sachin Krishnia[1], Johannes Güttinger[2], Armin Satz[2], Robert Frömter[1]*, and Mathias Kläui[1]

[1] Institute of Physics, Johannes Gutenberg University Mainz, 55099 Mainz, Germany

[2] Infineon Technologies Austria AG, 9500 Villach, Austria

*Electronic Mail: froemter@uni-mainz.de

**Abstract**

Magnetic skyrmions, topologically stabilized spin textures, are promising candidates for future memory devices and non-conventional computing applications due to their enhanced stability, non-linear interactions, and low-power manipulation capabilities. Despite their significant potential, the reliable electrical readout of individual skyrmions remains a fundamental challenge. While magnetic tunnel junctions and anomalous Hall effect (AHE)-based techniques have demonstrated skyrmion detection capabilities, they currently fail to reliably detect single moving skyrmions as required for applications. Our approach leverages thermally activated skyrmions, where a low constant drive current simultaneously generates both skyrmion motion and the Hall voltage necessary for detection. We demonstrate the reliability of this method through real-time correlation between measured Hall voltage signals and direct Kerr microscopy imaging. Two consecutive Hall crosses allow for determining the skyrmion velocity, in accordance to Kerr microscopy videos. We present an analytical formula for the skyrmion AHE readout signal, demonstrating scalability from micrometer to nanometer skyrmions. These advances establish a robust platform for skyrmion-based sensors, counters and unconventional computing systems that depend on precise individual skyrmion control and detection.

**Main Text**

Magnetic skyrmions are chiral magnetic whirls that exhibit enhanced stability due to their non-trivial topology[1–4]. This topological protection has generated considerable interest in skyrmions as potential candidates for next-generation computational applications[5–14] and high-density, energy-efficient data storage technologies[3,15–17] offering the potential for miniaturized low-power devices. Skyrmions are primarily stabilized by the competition between dipolar energy, Zeeman energy, exchange interaction, and Dzyaloshinskii-Moriya interaction. They can emerge both in bulk non-centrosymmetric magnets[18] and in thin magnetic multilayer[2,19–21], where they exhibit individual, particle-like behavior[3,5].

Current research explores two principal avenues for exploiting skyrmion properties in device applications. First, deterministic schemes utilize skyrmions as mobile information carriers[15], relying on controlled processes such as nucleation, annihilation, and motion to perform memory operations. Second, the stochastic and diffusive dynamics of skyrmions have opened pathways to unconventional computing concepts, as has been recently demonstrated by recognizing hand gestures,[22] and for Boolean logic-gate operations[23] based on thermally diffusive Brownian motion of skyrmions[8,13,24]. Further applications include reshuffling[5] and sensing devices[25], where the operational characteristics are directly linked to tunable skyrmion diffusion rates[8,26]. Furthermore, the diffusion can be enhanced by several orders of magnitude[26] for device integration. Both deterministic and stochastic approaches benefit from efficient manipulation of skyrmions via spin transfer torque[27,28] or spin-orbit torque[29,30] at ultra-low current densities. In thermally-activated systems, thermal excitations can overcome pinning effects, allowing directed motion of skyrmions at even lower current densities[5,23,31].

Despite the significant potential of skyrmions in spintronics device application, their electrical readout remains a major challenge to the date. Current strategies primarily involve magnetic tunnel junctions (MTJs)[32–34,34–39]. However, the MTJ readout has been demonstrated only for static skyrmions falling short of what is required for applications.

An alternative electrical read-out mechanism is using the Anomalous Hall effect (AHE)[40–43]. This has been recently demonstrated for ensembles of skyrmions, however, facing challenges due to the requirement of different current amplitudes for skyrmions motion and readout[44]. Moreover, this method applied to large areas only provides qualitative information about skyrmion numbers, due to non-identical skyrmion sizes, making precise counting inadequate. In another experiment, AHE-based readout of single static skyrmions was shown[41], but the electrical readout of dynamic skyrmions on-the-fly during motion has not yet been experimentally realized.

In this work, we investigate the dynamics of diffusive skyrmions in judiciously designed magnetic multilayers that require significantly reduced current densities to drive skyrmions through Hall crosses. Moving skyrmions are electrically detected at the Hall cross via the anomalous Hall effect using just their drive current to generate the Hall voltage, enabling real-time readout during their motion. This approach facilitates on-the-fly skyrmion detection and allows the use of narrow channels to count individual skyrmions. We validate this methodology through real-time synchronization with Kerr microscopy imaging. We further demonstrate the reliability of electrical detection by measuring skyrmion velocities between two Hall crosses, which shows excellent agreement with Kerr-based measurements. Our results establish a robust framework for electrical detection of moving skyrmions and open avenues for their integration in low-power sensors, memory devices and Brownian computing architectures.

Unless otherwise specified, all subsequent measurements were conducted on a Ta(4)/$Co_{20}Fe_{60}B_{20}$(0.9)/Ta(0.08)/MgO(2)/$HfO_2$(3) multilayer stack (nominal layer thicknesses in nanometers given in parentheses) deposited using a Singulus Rotaris magnetron sputtering system. The multilayers host skyrmions exhibiting spontaneous diffusion[22,23,26,45]. Skyrmions were stabilized by a small out-of-plane bias field of approximately $50\ \mu\mathrm{T}$. Skyrmion nucleation was achieved by in-plane field pulses of $30\ \mathrm{mT}$ at temperatures of $330-340\ \mathrm{K}$.

To detect the skyrmions via the AHE, we patterned the thin films into Hall bar devices. A schematic of the device with electrical measurement configuration is shown in Figure 1(a). The device comprises two skyrmion reservoirs which are connected through an 8 µm wide channel. Skyrmions with a typical diameter of approximately 3 µm are nucleated within these reservoirs. The width of the channel is chosen to ensure that only a single skyrmion can pass through the channel at a time. This selective transmission is warranted by skyrmion-skyrmion and skyrmion-edge repulsion effects[46]. When a current is applied between the reservoirs, skyrmions are driven via spin-orbit torques from one reservoir to the other through the Hall bar, generating measurable changes in the transverse voltage due to the AHE. We use these variations in the Hall voltage to detect the dynamics of individual skyrmions passing through the Hall bar. Here a current of 10 µA, corresponding to a current density of $2.9\times10^8$ A/$m^2$, is applied.

Due to the micrometer-scale size of the investigated skyrmions, in parallel we employ non-perturbative Kerr microscopy to capture their current-induced dynamics in real time and correlate these observations with the measured Hall voltage. This approach enables a one-to-one correlation between Hall voltage variations and skyrmion passage, confirming the single-skyrmion sensitivity of the skyrmion sensor. Figure 1 demonstrates the passage of individual skyrmions through the Hall bar and the corresponding change in the Hall voltage.

Figure 1(b) shows the Hall voltage as a function of time where each voltage peak corresponds to the passage of a single skyrmion from the right to the left reservoir. To verify that the observed peaks in the Hall voltage indeed correspond to skyrmion transits, we analyze a representative event in Figure 1(c), correlating the Hall signal with Kerr microscopy. At $t$ = 10.3 s, no skyrmion is present at the Hall bar and the measured Hall voltage is at baseline $U$ = 19.1 µV. As a skyrmion enters the Hall cross region at $t$ = 10.5 s, the voltage rises to $U$ = 22.5 µV. The corresponding Kerr image confirms the presence of a skyrmion at the center of the Hall cross. Subsequently, at $t$ = 10.7 s, the Hall voltage drops to baseline again, indicating that the skyrmion has exited the sensitive region. The skyrmion velocity, estimated from time-resolved Kerr images, is approximately 50 µm/s. Under these conditions, the Hall signal of the full hysteresis loop is 50 µV. The reduced signal observed in the skyrmion measurements is due to the smaller size of the skyrmions compared to the channel size. To validate the measured difference in Hall voltage, we calculated the passage of a single skyrmion through the cross. The skyrmion size is estimated from Kerr microscopy images, while the resistivity tensor is determined based on the device geometry and experimentally measured longitudinal and Hall resistances. Using these parameters, we simulate and analytically calculate the expected Hall voltage generated by a single skyrmion, see Figure 1(d). For narrow Hall contacts and small anomalous Hall angle $\rho_{xy}/\rho_{xx}$ the skyrmion peak signal $\Delta R_{xy}$ is given by

$$\Delta R_{xy} = \frac{\rho_{xy}}{t}\frac{2\mathcal{A}(R)}{W^2}, \tag{1}$$

where $t$ is the thickness, $\rho_{xx}$ and $\rho_{xy}$ are the longitudinal and transversal resistivity, respectively, $\rho_{xy}/t$ is the Hall signal of the saturated channel, $W$ the channel width, and $\mathcal{A}(R)$ the effective area of the skyrmion, which for negligible skyrmion wall width (compared to the diameter) is simply $\mathcal{A}(R) = \pi R^2$. This expression results in a constant signal as long as the channel width $W$ is scaled proportionally to the skyrmion size $R$. The formula is derived in the Supplementary Material, where also the more general cases of wide contacts and arbitrary skyrmion position are treated. The calculated voltage differences are consistent with those observed experimentally.

Under dark conditions (no illumination), the measured noise of the anomalous Hall signal approaches the thermal noise limit, which is given by $U = \sqrt{4k_B T R \Delta f} \approx 60\ \mathrm{nV}$ with the transverse resistance $R \approx 4\ \mathrm{k\Omega}$ and the bandwidth $\Delta f = 50\ \mathrm{Hz}$. However, when the sample is illuminated during Kerr microscopy measurements, the noise level increases significantly, due to photovoltage generated in the sample. This effect also adds a constant offset to the measured voltage. The effect is highly dependent on the light intensity and the specific device. To minimize the photovoltage, minimum light is used to illuminate the devices. In an all-electrical device this effect will not occur. Additionally, a capping layer of $HfO_2$ is added to enhance the observed Kerr microscopy contrast. Besides the optical properties of $HfO_2$, the insulating characteristics result in less voltage being shunted, improving signal to noise ratio in AHE.

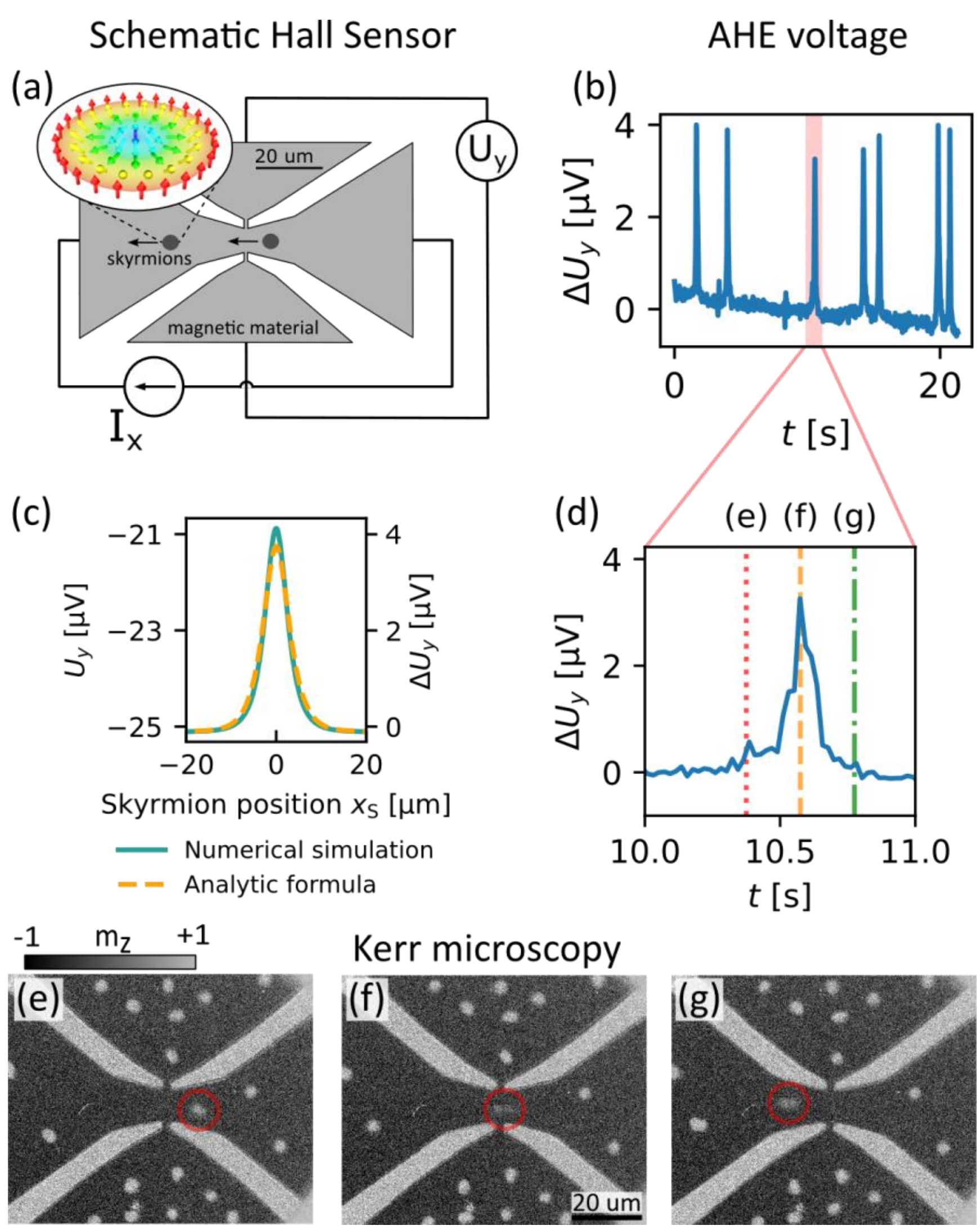

**Figure 1: Electrical detection and Kerr imaging of skyrmions in Hall bar.** (a) Schematic of the Hall bar with the magnetic stack shown in gray and skyrmions in black. The arrows represent the direction of motion of skyrmions under applied current. A schematic spin configuration of a skyrmion is shown in the inset. (b, d) Measured Hall voltage as a function of time. (c) Numerical simulation and analytical of the AHE signal; the deviation arises from the non-straight channel and the finite Hall contact width, which are geometrical factors outside the scope of the analytical model. (d)

Shows a zoom into the Hall voltage during passage of a single skyrmion through the Hall cross. (e-g) Kerr microscopy images corresponding to the times (e) to (g) marked in panel (d). At the four light grey lines all metallic layers have been removed by etching to guide current, voltage and skyrmions. The magnetic skyrmions appear as bright dots in black background in the differential Kerr images. The skyrmions inside the Hall contacts (top and bottom) do not contribute to the measured Hall signal, as no current flows in these parts. The moving skyrmions appear smeared out or even doubled due to the finite exposure time of Kerr microscope (65 ms). The apparent constant baseline drift of the Hall voltage in (b) is an artifact resulting from a disabled drift correction to increase the data rate of the used nanovoltmeter Keithley 2182.

Although our measurements demonstrate that moving skyrmions can be detected in the Hall bar via AHE, it remains unclear whether all skyrmions are reliably detected or not and no skyrmions are annihilated or nucleated in the Hall bar, a critical concern for device applications. To assess the detection fidelity of the AHE-based readout mechanism, we next fixed the total the number of skyrmions by confining the reservoir area. For simultaneous electrical and optical detection, the device was positioned within the field of view of the Kerr microscope. Both skyrmion reservoirs were spatially confined using low-dose milling ($5 \times 10^{-15}\,\frac{\mathrm{C}}{\mu\mathrm{m}^2}$) with a 30 keV focused Ga ion beam, by etching 1 μm wide bounding lines shown as red vertical lines in Figure 2(a). This process locally reduces the magnetic anisotropy[47–51], creating boundaries that repel magnetic textures such as skyrmions, while preserving full electrical conductivity[47]. The reduced reservoir dimensions enable visualization of the entire device within a single field of view using Kerr microscopy, as no skyrmions can escape sideways into the current leads.

In Figure 2, we show a reliable electrical detection of skyrmions using the AHE. We nucleate four skyrmions in the right reservoir as shown in Figure 2(a) and subsequently drive them into the left reservoir by electrical current and detect them using AHE. The time dependent Hall voltage exhibits four discrete peaks, each corresponding to the transit of an individual skyrmion through the detection region. The corresponding Kerr image in Figure 2(b) verifies their transfer into the left reservoir. We then reverse the direction of current and drive the skyrmions back in Figure 2(c), from left to right, to verify the bidirectional motion and the reproducibility of the AHE-based detection. Notably, the Hall voltage traces in both directions exhibit four distinct peaks, confirming that the number of skyrmions remains unchanged during transit, and that neither annihilation nor nucleation occurs during the readout process. These results establish the fundamental reliability of AHE-based detection in counting multiple skyrmions. In total, twelve individual transits of four skyrmions across the Hall bar consistently recorded four discrete signals per cycle, confirming highly reproducible detection with no counting errors.

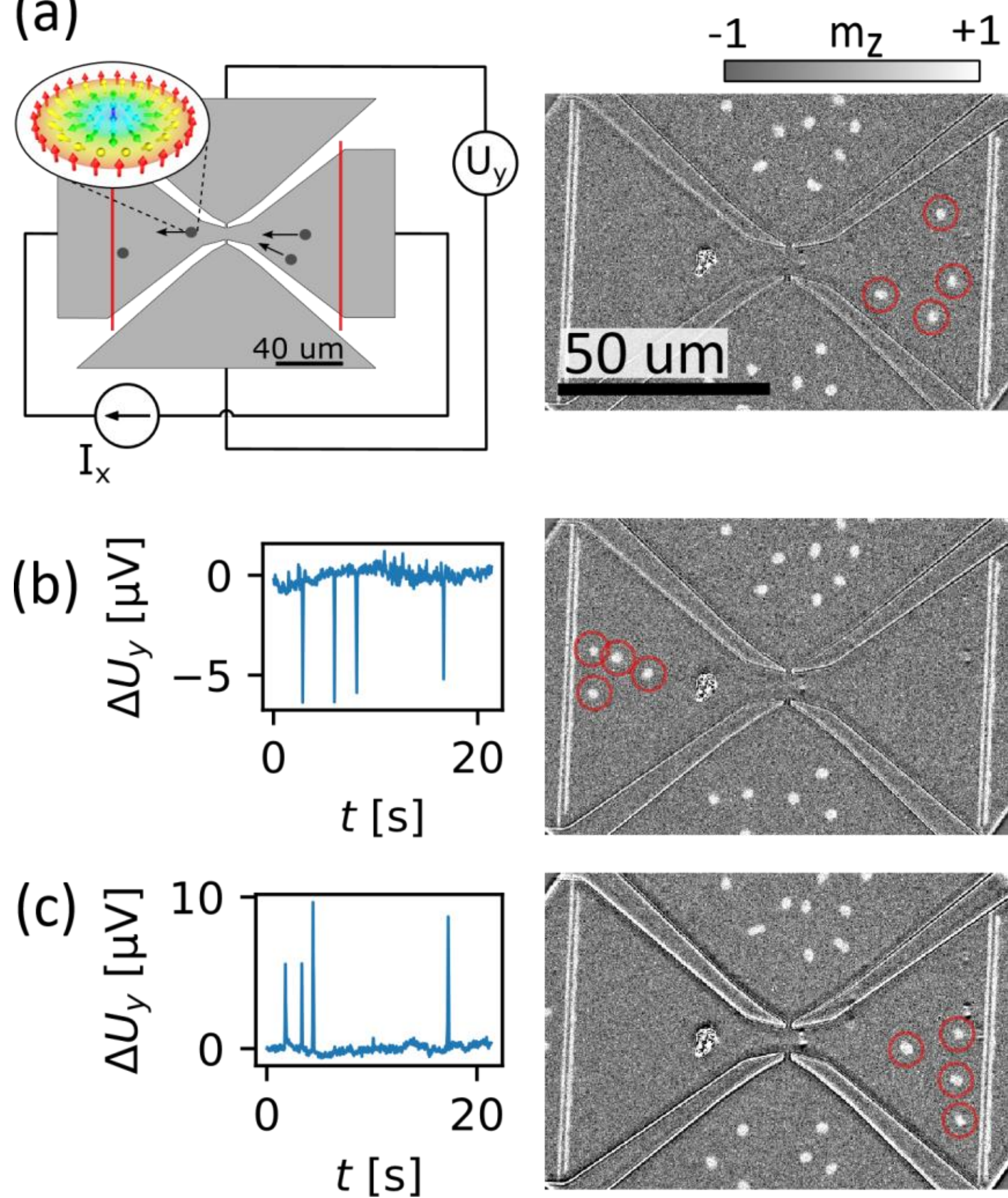

**Figure 2: Electrical detection of skyrmions in the irradiated skyrmion Hall bar device.** (a) Schematic representation showing the magnetic layer in grey and the two FIB-irradiated lines in red with an area dose of $5 \times 10^{-15} \frac{\mathrm{C}}{\mu\mathrm{m}^2}$ and a width of 1 µm. Additionally we show a Kerr microscopy image with four skyrmions (circled by red) before currents are applied. (b, c) Time-dependent Hall voltage measurements with corresponding Kerr microscopy images taken after each measurement. The relevant skyrmions, which are moved through the Hall bar, are marked with red circles.

Beyond skyrmion counting, an additional experimental challenge lies in measuring the velocity of skyrmions for racetrack memory and neuromorphic computing applications. This requires real-time electrical detection of moving skyrmions at multiple spatial positions along a track. To date, most electrical studies have focused on static detection schemes—i.e., skyrmions are first driven and subsequently detected using static imaging techniques or via AHE[44]. However, real-time electrical detection of moving skyrmions at multiple spatial locations remains to be shown.

To address this, we fabricated a dual Hall cross geometry, as illustrated in Figure 3(a). In addition to employing a single Hall cross for skyrmion detection, the use of multiple crosses enables spatiotemporal resolution of the skyrmion motion. This configuration enables the extraction of directionality and velocity through correlation analysis. Figure 3(c) shows the variation in Hall voltage measured at Hall cross 1 (in blue) and 2 (in orange). The time delay between these signals enables determination of individual skyrmion velocities as shown in Figure 3(b) of 400 skyrmion transits. We compute this time difference by a threshold-based algorithm: a skyrmion is considered to have arrived at a Hall cross when the signal exceeds a predefined threshold (threshold is shown as green line in Figure 3(c). The velocity $v_{\mathrm{AHE}} = L/\Delta t$ is calculated with the sensor spacing $L = 20\ \mu\mathrm{m}$ and the time difference between arrival times $\Delta t$. This method is selected for its simplicity in future all-electrical implementations. A representative event of a single skyrmion passing is shown in Figure 3(d). The skyrmion velocity exhibits fluctuations due to the stochastic nature of thermally assisted current driven motion at these current densities. At an applied current of 18 mA ($3\times10^8$ A/m$^2$), the average velocity of skyrmions extracted from the dual Hall voltage measurement is 151(61) µm/s. Besides the Hall measurements, the simultaneous Kerr microscopy measurements are also evaluated. Due to the limited time resolution of only 16 Hz a single skyrmion passing only consists of only 1-5 frames. Nevertheless, for the events shown in Figure 3(c), the skyrmion velocities extracted from Kerr imaging $\bar{v}_{Kerr} = 130(69)\ \mu\text{m/s}$, are in full agreement with those obtained from the dual AHE readout $\bar{v}_{AHE} = 136(20)\ \mu\text{m/s}$.

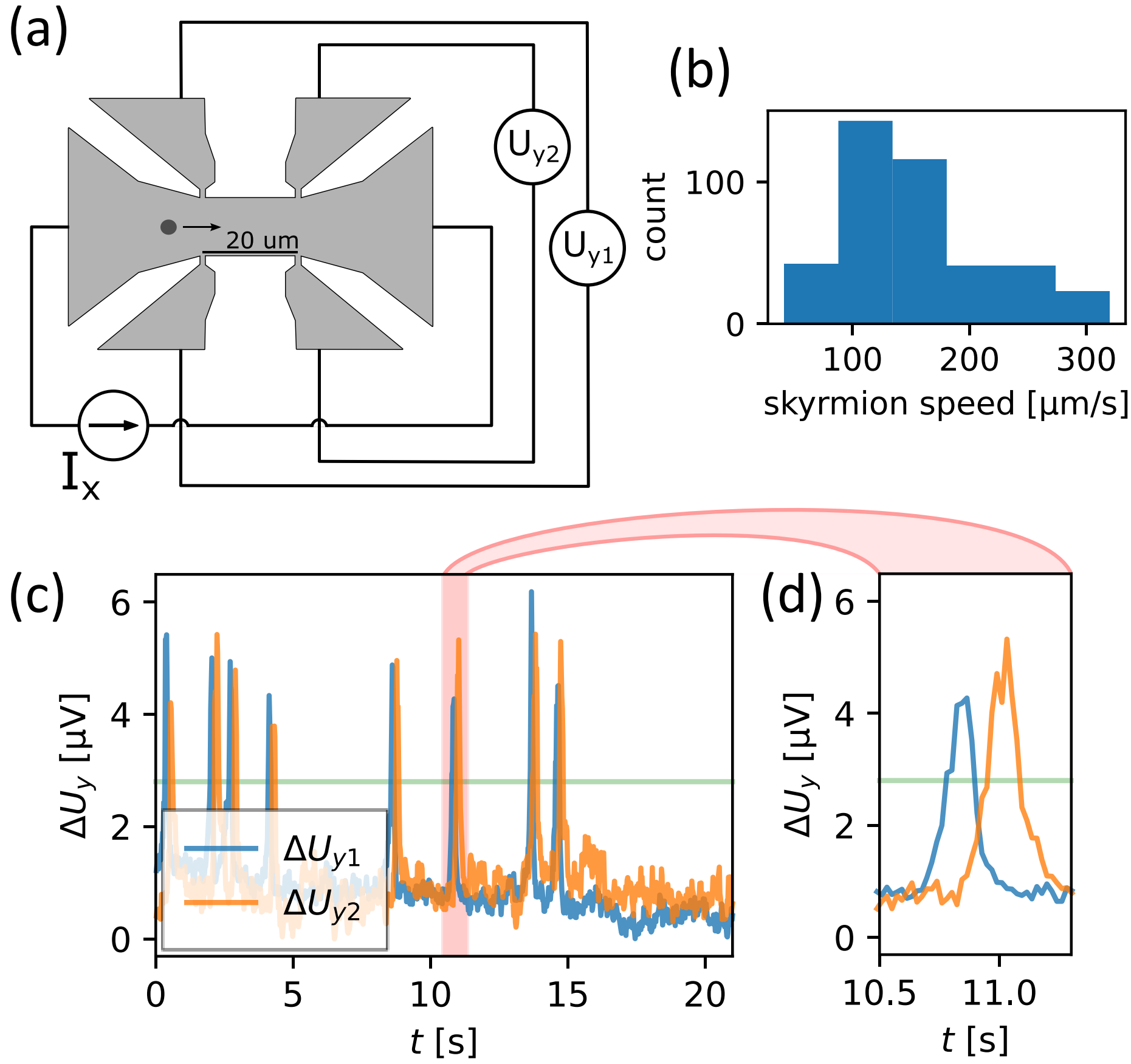

**Figure 3: Electrical detection of skyrmions using a dual Hall cross.** (a) Schematic of the dual Hall cross device. (b) Computed skyrmion velocities out of the sensor spacing and the time difference in Hall signals from (c), 400 Skyrmion passings are measured in total. (c,d) Measured Hall voltages from double Hall cross showing the temporal evolution. Skyrmions are detected with one threshold shown as a green line. If the signal surpasses the green threshold it counts as the passing of a skyrmion. (d) Detailed view of a skyrmion passing in (c).

Finally, we discuss and demonstrate strategies to enhance the signal-to-noise ratio in the electrical readout of moving skyrmions. We employ a double-repetition sample with the stack configuration: Ta(4)[Ta(2)$Co_{20}Fe_{60}B_{20}$(0.9)Ta(0.04)MgO(2)]$_{\times 2}$HfOx(4). The skyrmions are nucleated with an in-plane (IP) field pulse of 90 mT and stabilized by a constant bias OOP field of approximately 300 µT at room temperature. The inclusion of two ferromagnetic layers increases the magnetic volume, thereby enhancing thermal stability. However, it also necessitates higher current densities to drive skyrmions, owing to the increased magnetic volume, additional current shunting paths, and the introduction of more pinning sites. Furthermore, increasing the number of multilayer repetitions leads to a reduction in skyrmion size[52], making them difficult to resolve using Kerr microscopy. Consequently, we limit our investigation to stacks with two repetitions, while more repetitions can still be detected electrically.

To drive the skyrmions, a current of 300 µA ($4.3\times10^9$ A/m$^2$) is required, resulting in a Hall signal for the fully saturated channel of 1 mV. Additionally, an out-of-plane (OOP) field oscillation (300 µT at 50 Hz) is applied to mitigate pinning effects[26]. Due to the increased current required for the skyrmion motion, the signal-to-noise ratio of the measured Hall voltage is significantly enhanced by a factor of 5 compared to Figure 2. While this improvement is lower than anticipated based solely on current magnitude, it is consistent when considering the reduction of the skyrmion size (see Supplementary Material for more details). As demonstrated in Figure 4, the improved signal-to-noise ratio enables the detection of skyrmions positioned further away from the Hall cross. Notably, skyrmions located at the positions shown in Figure 4(c) and (e) still result in a sizeable Hall voltage as can be seen in Figure 4(b).

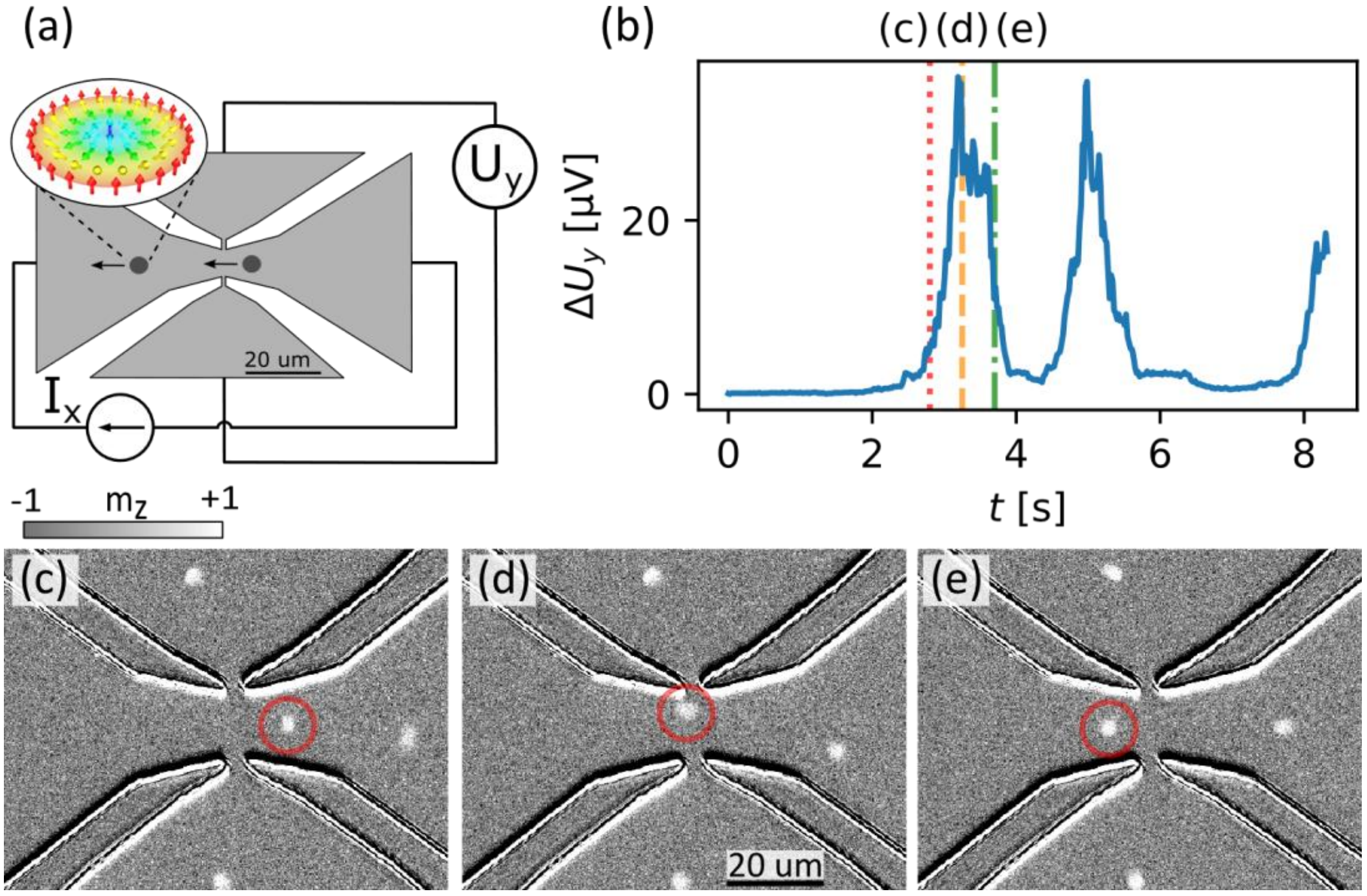

**Figure 4: Hall signature of skyrmions in the double-layer stack.** (a) Schematic of the Hall bar device with magnetic stack shown in grey. (b) Temporal evolution of the measured Hall voltage with skyrmions passing the cross. (c-e) Kerr microscopy images captured at points in time (c) to (e) as marked in (b), showing skyrmion motion through the device. Due to thermal drift the background subtraction becomes visible in the Kerr image, showing the contours of the confining structure.

In conclusion we have demonstrated reliable detection and counting of individual skyrmions using the AHE, complemented by real-time Kerr microscopy to directly correlate optical and electrical signals. We established quantitative agreement between the single-skyrmion signature measured by AHE and a current-density distribution model which can easily be extended to different geometries. We found that the same skyrmions can be passed at least than ten times back and forth through the sensor with perfect counting ability and no annihilation or nucleation. By implementing dual Hall cross devices, we achieved improved readout precision and simultaneously measured the skyrmion velocities for over 400 skyrmions. These velocity measurements are in excellent agreement with the simultaneous velocity measurements from Kerr microscopy data. Furthermore, we explored a multilayer stack configuration consisting of two identical dipolar-coupled ferromagnetic layers. This configuration substantially enhances skyrmion stability and enhances the AHE signal to noise ratio. Although we limited our study to single- and double-repetition stacks, further optimization of layer composition or the adoption of higher-repetition architectures could broaden the operational field range and further improve the measured noise. Besides, utilizing stacks with smaller, non-optically detectable skyrmions represents another promising approach for enhancing the operating field range. Equation (1) predicts that the AHE signal of a skyrmion is independent of its size, provided the channel dimensions are scaled proportionally and will even slightly increase for nanoscale skyrmions. Smaller skyrmions could be obtained, e.g., by changing materials or increasing the number of magnetic repetitions in the stack[44,53], thus also the AHE detection of nanoscale skyrmions in racetrack devices will be feasible.

**Supplementary Material**

In the supplemental material, an analytical model for the AHE signal from a skyrmion in a Hall bar geometry is derived and validated by simulation. We include the cases of finite skyrmion wall width, as well as finite widths of the Hall contacts. In addition, the sample preparation is described, and a hysteresis curve of the multilayer is given.

**Declaration of Interest**

The authors have no conflicts to disclose.

**Author Contributions**

G.B. performed the experiments in cooperation with N.V. and wrote the manuscript with K.L., N.V, F.K., D.M.T, S.K., J.G., R.F. and M.K.. K.L. performed supporting simulations and provided the analytical formulas. F.K. and D.M.T. grew the materials stacks used in the experiments. J.G., A.S., R.F. and M.K. supervised the project.

**Data availability**

Data that supports the findings of this study will be openly available in Zenodo upon acceptance of this manuscript at https://doi.org/10.5281/zenodo.17464619.

**Acknowledgements**

The work in Mainz (G.B., K.L., F.K., M.D.T., S.K., R.F. and M.K.) was supported by the Deutsche Forschungsgemeinschaft (DFG, German Research Foundation) projects 403502522 (SPP 2137 Skyrmionics), 49741853, and 268565370 (SFB TRR173 projects A01, B02 and A12) as well as TopDyn and the Zeiss foundation through the Center for Emergent Algorithmic Intelligence. This research was supported by the National Research Council of Science & Technology (NST) grant by the Korean government MSIT (Grant No. GTL24041-000). The work is a highly interactive collaboration supported by the Horizon 2020 Framework Program of the European Commission under FET-Open grant agreement no. 863155 (s-Nebula) and ERC-2019-SyG no. 856538 (3D MAGiC) and the Horizon Europe project no. 101070290 (NIMFEIA), which M.K. acknowledges. This project has received funding from the European Union's Horizon Europe Programme Horizon.1.2 under the Marie Skłodowska-Curie Actions (MSCA), Grant agreement No. 101119608 (TOPOCOM). This work was developed in collaboration with and supported by Infineon Technologies Austria AG.